\documentclass[sigconf, nonacm]{acmart}

\usepackage{multirow}
\usepackage{tabularx}

\AtBeginDocument{%
  \providecommand\BibTeX{{%
    \normalfont B\kern-0.5em{\scshape i\kern-0.25em b}\kern-0.8em\TeX}}}

\begin{document}

\title{Self-supervised Human Activity Recognition by Learning to Predict Cross-Dimensional Motion}


\author{Setareh Rahimi Taghanaki, Michael Rainbow, Ali Etemad} 
\affiliation{%
  \institution{Queen's University}
  \streetaddress{99 University Avenue}
  \city{~Kingston}
  \state{ON}
  \country{Canada}
}
\email{{19srt3, michael.rainbow, ali.etemad}@queensu.ca }





\begin{abstract}
We propose the use of self-supervised learning for human activity recognition with smartphone accelerometer data. Our proposed solution consists of two steps. First, the representations of unlabeled input signals are learned by training a deep convolutional neural network to predict a segment of accelerometer values. Our model exploits a novel scheme to leverage past and present motion in \textit{x} and \textit{y} dimensions, as well as past values of the \textit{z} axis to predict values in the \textit{z} dimension. This cross-dimensional prediction approach results in effective pretext training with which our model learns to extract strong representations. Next, we freeze the convolution blocks and transfer the weights to our downstream network aimed at human activity recognition. For this task, we add a number of fully connected layers to the end of the frozen network and train the added layers with labeled accelerometer signals to learn to classify human activities. We evaluate the performance of our method on three publicly available human activity datasets: UCI HAR, MotionSense, and HAPT. 
The results show that our approach outperforms the existing methods and sets new state-of-the-art results.\end{abstract}


\keywords{Self-supervised Learning, Accelerometer Data, Human Activity Recognition, Data Prediction, Machine Learning.}


\maketitle

\vspace*{0.75cm}
\section{Introduction}
\label{sec:intro}

Human activity recognition (HAR) has become a popular field of research due to its wide applications in health \cite{osmani2008human, oniga2014human}, personalization of user experience \cite{zhao2017keeping}, sports \cite{direkoglu2012team, ermes2008detection}, and others. While activity recognition can be carried out using data from a variety of sources such as videos or motion capture data, the pervasiveness of inertial measurement units (IMU) in smartphones and wearable devices \cite{wang2016comparative, kulchyk2019activity} has created new research opportunities for HAR studies. Moreover, advances in machine learning and deep learning in recent years has resulted in a variety of different solutions being proposed for HAR \cite{saeed2019multi, wang2016comparative}.

\let\thefootnote\relax\footnote{https://doi.org/10.1145/3460421.3480417}

Existing HAR solutions often rely on fully supervised techniques \cite{oniga2014human, wang2016comparative}. While these approaches are generally robust and effective in terms of performance \cite{oniga2014human}, they face a number of challenges and limitations. First, since the amount of training data highly influences the performance of deep neural networks, large \textit{labeled} datasets are required to train activity recognition models. Due to the increasing use of smartphones in daily life, massive amounts of wearable inertial data has been collected. Nonetheless, annotating such large time-series datasets is generally expensive and time consuming. Furthermore, fully-supervised models are usually trained from scratch for every task, which not only is time consuming and computationally expensive, but also results in learning very task-specific representations that cannot be easily used for other similar tasks. This also makes the aggregation of multiple datasets difficult given the various output classes (labels) available in different datasets.

In this paper we propose a \textit{self-supervised} framework for HAR with the goal of reducing reliance on labeled data while achieving a strong performance. Our proposed solution uses two separate steps, namely pretext representation learning and downstream activity recognition. The self-supervised pretext task focuses on \textit{cross-dimensional motion prediction}. Specifically, we use the past values of the \textit{z} IMU axis, as well as past and present values of \textit{x} and \textit{y} to predict the values of \textit{z}. While motion prediction has been used in the field of computer vision for self-supervised learning \cite{srivastava2015unsupervised, wang2019self}, our model is novel in that the cross-dimensional aspect results in more effective learning.
As a result of the self-supervised representation learning, the proposed model learns generalized accelerometer features that can be used in motion-related tasks such as HAR across different datasets. The model is then frozen and transfer learning is used to perform activity recognition. Our contributions can be summarized as follows. (\textbf{1}) We propose a novel self-supervised solution for HAR based on cross-dimensional motion prediction. This is the first time that motion prediction with cross-dimensional learning has been used for self-supervised learning in the context of HAR. We demonstrate that using our solution to learn the pretext task, generalized and informative motion representations are learned. 
(\textbf{2}) We test our method on \textit{three publicly available datasets}, UCI HAR \cite{anguita2013public}, MotionSense \cite{malekzadeh2018protecting}, and HAPT \cite{reyes2016transition}. Our results outperform other methods for these datasets and sets new \textit{state-of-the-art}.

\section{Related Work}
\label{sec:relatedWork}
\subsection{HAR with Fully Supervised Learning}
A variety of different data modalities have been used in the past for HAR, for example, vision \cite{babiker2017automated}, IMU \cite{kulchyk2019activity, yao2018efficient,yu2018multi}, motion capture \cite{ijjina2014one}, and even audio \cite{chahuara2016line}, among others. Given the ubiquity of IMUs in smartphones and wearables, in our work, we focus on accelerometer-based HAR, which has been explored in recent years using different machine learning and deep learning methods \cite{kulchyk2019activity, yao2018efficient,yu2018multi}.
\begin{figure*}[t!]
\centerline{\includegraphics[width=1\linewidth]{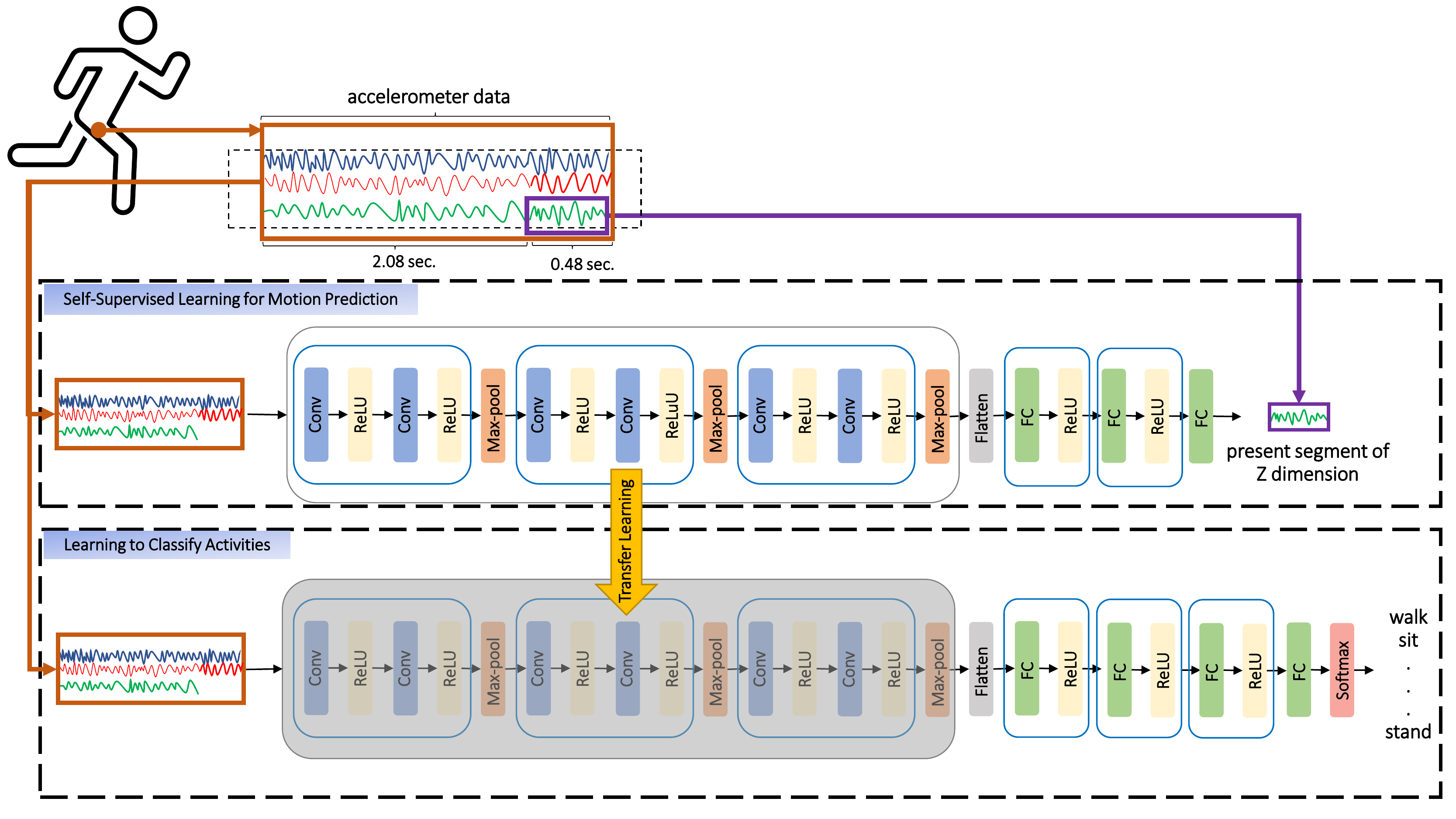}}
\caption{The proposed self-supervised architecture is presented. } 
\label{fig:model architecture}
\end{figure*}

Convolutional neural networks (CNN) have been widely used for IMU-based HAR \cite{kulchyk2019activity, yao2018efficient, bianchi2019iot}. For example, in \cite{kulchyk2019activity}, a simple 5-layer CNN was used to perform activity recognition and analyze the impact of IMUs on different parts of the body. Similarly, a 5-layer CNN was used in \cite{yao2018efficient} to perform labeling of activity data with variable-sized windows. In \cite{bianchi2019iot}, a relatively deeper CNN was used to perform activity recognition in smart homes. Recurrent neural networks (RNN), either alone or accompanying CNNs, have also been widely used for HAR applications \cite{yu2018multi, yu2018human, xia2020lstm}. In \cite{yu2018multi}, LSTM cells were useed to implement a multilayer parallel RNN model for data collected from a waist-worn smartphone. In \cite{yu2018human}, a bi-directional LSTM network was used to process both the past and future time-steps of the input signals. In \cite{xia2020lstm}, a deep neural network combining 2 convolution and 2 LSTM layers was presented for the purpose of HAR with smartphone data. A detailed survey on different deep learning solutions for HAR is available in \cite{nweke2018deep}.

\subsection{HAR with Self-Supervised Learning}
Although supervised deep learning models have shown significant performance for HAR, to reduce the reliance on labeled data, semi-supervised and self-supervised learning methods have been explored in recent years. Self-supervised learning has been previously explored in time-series representation learning applications such as computer vision \cite{lee2017unsupervised, jing2018self}, bio-signal analysis (e.g. ECG \cite{sarkar2020self, sarkar2020self2} and EEG \cite{banville2019self}), speech recognition \cite{ravanelli2020multi}, and others. For IMU-based HAR, \cite{saeed2019multi} and \cite{haresamudram2020masked} have used self-supervised learning. In \cite{saeed2019multi}, 6 transformations were applied to unlabeled IMU data, and a multitask deep neural network was used to recognize the transformations as the pretext self-supervised tasks. In \cite{haresamudram2020masked}, certain data-points were masked and reconstructed as the pretext self-supervised task. Both models were then frozen and used as the feature extractor for downstream activity recognition. Our work differs from \cite{haresamudram2020masked} in that \cite{haresamudram2020masked} masks only \textit{individual} samples throughout the accelerometer time-series, while ours masks entire segments consisting of consecutive samples. Moreover, our model uses cross-dimensional learning to leverage some axis to predict motion in the other axis, which results in more effective pre-text learning. In \cite{tang2021selfhar}, teacher-student self-training and multi-task self-supervision were combined to perform HAR task, in the context of semi-supervised learning.

\section{Method}
\label{sec:method}

\subsection{Proposed Self-supervised Architecture}
\label{ssec:arch}

The goal of this study is to develop a self-supervised deep neural network capable of learning robust and generalizable representations from accelerometer signals for human activity recognition. To this end, we train a network $H_{predict}$ in a self-supervised manner using pretext task $T_{pretext}$, where this stage would allow $H_{predict}$ to learn representations capable of performing a downstream human activity recognition task $T_{downstream}$. We set $T_{pretext}$ to a problem in which given $\Phi_{input} = [[\phi^{x}_1, \phi^{x}_2, \ldots, \phi^{x}_{r+h}]^T, [\phi^{y}_1, \phi^{y}_2, \ldots, \phi^{y}_{r+h}]^T, $ \\$ [\phi^{z}_1, \phi^{z}_2, \ldots, \phi^{z}_r]^T]$ where $\phi^x_i$ is the $i$th time-step of the motion sequence in the \textit{x} axis. The goal of our network is to predict $\Phi_{predict} = [\phi^z_{r+1}, \ldots, \phi^z_{r+h}]$. In essence, the $H_{predict}$ uses past and present values of \textit{x} and \textit{y}, along with past values of \textit{z} to predict a present segment of \textit{z}, i.e., $\Phi_{predict}$. Next, the goal is to use a model $H_{HAR}$, consisting of model $H_{predict}$ frozen and augmented with a few additional fully connected (FC) layers, to learn $T_{downstream}$ defined as the activity recognition task. Figure \ref{fig:model architecture} presents an overview of our proposed self-supervised pipeline. Through the following, we describe $H_{predict}$ and $H_{HAR}$, the loss functions used, and the implementation details.

We use a deep CNN as our $H_{predict}$ to learn motion representations through $T_{pretext}$. The CNN consists of 3 convolutional blocks each of which contains two convolution + ReLU layers. Each block is followed by a max-pool layer. This structure is then followed by a flatten operation to obtain a 1D tensor. Next, FC layers accompanied by ReLU activation functions are used to learn the non-linearities between the learned representations and the outputs. Two such layers yield the best results for our application and are therefore used. Finally, the output layer consists of a single FC layer. MSE loss is used to train the network. 
The network is fed motion signals of 2.56 seconds from axes \textit{x} and \textit{y}, and 2.08 seconds of axis \textit{z}, while the subsequent 0.48-second segment of axis \textit{z} is used as the output.
Each input has a 50\% overlap with the next input segment. The data in each dataset are normalized with the min-max method. Once $H_{predict}$ has been trained, we freeze the convolution blocks and transfer the frozen weights to $H_{HAR}$. 
The network is then augmented with $3$ additional FC + ReLU blocks. Lastly, a single FC layer followed by sigmoid activation is added as the output layer. $H_{HAR}$ is then trained while the convolution blocks are kept frozen and act as feature extractors for the HAR network. Entropy loss is used to train the network and 20\% dropout is set for the 1$^{st}$ FC block.
The number and types of blocks/layers (for both networks), as well as the segment sizes used for training, are all tuned empirically to obtain optimum performance. Figure \ref{fig:model architecture} presents the architecture of the self-supervised model used for motion representation learning through predicting future motion values.

\subsection{Implementation Details}
\label{ssec:implementation}

In order to implement our models, we used Keras with TensorFlow backend on an NVIDIA GeForce RTX 2080 Ti GPU. The $H_{predict}$ kernel size is set to 3 for all the convolution layers. The number of filters is set to 128, 256, and 384 for the three convolution block respectively, and the number of hidden neurons in the FC layers is set to 384, 120, and 24 for the three layers respectively. For the activity recognition network $H_{HAR}$, the frozen network is followed by 3 FC layers with 512, 250, and 100 neurons respectively, followed by a layer with the size of the number of activity classes.

$H_{predict}$ is trained over 80 epochs with a batch size 512 and Adam optimizer (learning rate $= 0.0003$). For $H_{HAR}$, a batch size of 512 and Adam optimizer (learning rate $= 0.0001$), are used for all three datasets. All the hyper-parameters of $H_{predict}$ and $H_{HAR}$ are summarized in Table \ref{tab:SSL_arch}.

In order to evaluate our method, we use the user-split based hold-out validation, randomly picking 20\% of the participants from each dataset for testing, while using the rest for training. Furthermore, 10\% of the training set is allocated for validation. We repeat this split scheme 5 times in a way that all users have been used once in the hold-out test set and 4 times in the training set. In the end, we average the results obtained by all models and present them as our method's results. The performance of $H_{predict}$ is evaluated with $R^2$ since it is a regression task, while $H_{HAR}$ is evaluated with accuracy and F1 metrics (mean and weighted). 

\section{Experiments and Results}
\label{sec:expRes}

\subsection{Datasets}
\label{ssec:Datasets}
Three publicly available datasets, UCI HAR \cite{anguita2013public}, MotionSense \cite{malekzadeh2018protecting}, and HAPT \cite{reyes2016transition} are used to evaluate our model. These datasets contain a variety of human activities and have been collected from smartphone sensors. Below, the characteristics of each dataset are briefly described.

\begin{table}[]
    \footnotesize
    \centering
    \caption{Architecture and parameters of the proposed model is presented.}
    \begin{tabular}{l|l|l}
    \hline
     \textbf{Module} & \textbf{Layer Details} & \textbf{Feature Shape} \\ \hline \hline
      Input & $-$ & $120 \times 3$ \\ \hline
      \multirow{7}{*}{Conv. Blocks} & $[\textit{conv}, 1 \times 3, 128] \times 2$ &   $116 \times 128$  \\ 
                                     &  $[\textit{maxpool}, 1 \times 2, \textit{stride} = 2]$ &   $58 \times 128$  \\ 
                                     &  $[\textit{conv}, 1 \times 3, 256] \times 2$ &   $54 \times 256$  \\ 
                                     &  $[\textit{maxpool}, 1 \times 2, \textit{stride} = 2]$ &   $27 \times 256$  \\ 
                                     &  $[\textit{conv}, 1 \times 2, 384] \times 2$ &   $23 \times 384$ \\
                                     &  $[\textit{maxpool}, 1 \times 2, \textit{stride} = 2]$ &   $11 \times 384$  \\ 
                                     & \textit{flatten} &  $4224
$ \\ \hline 
     \multirow{4}{*}{$H_{predict}$ FC} &  $[\textit{dense}, 384]$ &   $384$  \\ 
                                            &  $[\textit{dense}, 120]$ &   $120$  \\ 
                                            &  $[\textit{dense}, 24]$ &   $45$  \\
                                            & \textit{output} &  $24
$ \\ \hline 
     \multirow{7}{*}{$H_{HAR}$ FC} &  $[\textit{dense}, 512 ] $ &   $512$  \\ 
                                            &  $[\textit{dense}, 250]$ &   $250$  \\
                                            &  $[\textit{dense}, 100]$ &   $100$  \\
                                            &  $[\textit{dense}, num\_classes]$ &   $num\_classes$  \\
                                            & \textit{output} &  1
 \\ \hline 

    \end{tabular}
    \label{tab:SSL_arch}
\end{table}

\subsubsection{UCI HAR}
\label{ssec:data_UCI HAR}
This dataset contains accelerometer and gyroscope signals from 30 participants in the 9-48 age range (sampling rate of 50 \textit{Hz}). Each participant has worn a Samsung Galaxy S2 smartphone around their waist while performing 6 activities (walking downstairs, walking upstairs, walking, sitting, standing, and laying down).

\subsubsection{MotionSense}
\label{ssec:data_MotionSense}
This dataset consists of accelerometer, gyroscope, and altitude data collected with an iPhone 6s using SensingKit with a sampling rate of 50 \textit{Hz}. 24 participants of different gender, age, weight, and height groups have performed 6 activities (walking downstairs, walking upstairs, walking, sitting, standing, and jogging) in 15 trials, while carrying the smartphone in their trousers’ front pocket.

\subsubsection{HAPT}
\label{ssec:data_HAPT}
This dataset consists of 12 activities (6 basic activities: walking, walking downstairs, walking upstairs, standing, sitting, and lying, and 6 postural transitions: stand-to-sit, sit-to-stand, sit-to-lie, lie-to-sit, stand-to-lie, and lie-to-stand), collected from 30 participants with Samsung Galaxy S2 devices worn on their waists. Accelerometer and gyroscope signals have been captured with a sampling rate of 50 \textit{Hz}).

\subsection{Performance and Comparison}
\label{sec:majhead}

First, we evaluate our self-supervised network trained for the pretext task of motion prediction by measuring $R^2$ values for the predicted motion vs. the ground truth values. We obtain $R^2 = 0.682 \pm 0.062$ indicating strongly learned representations for motion prediction. 

Next, we evaluate the performance of our model for the downstream task of HAR ($H_{HAR}$). 
Tables \ref{tab:UCI_score}, \ref{tab:MotionSenhse_score}, and \ref{tab:HAPT_score} present the accuracy and F1 scores (mean and weighted) for the UCI HAR, MotionSense, and HAPT datasets, respectively, in comparison to other self-supervised methods in the literature that have also used these three datasets. 
The results illustrate that our method outperforms prior self-supervised methods in the field and sets a new state-of-the-art for UCI HAR, MotionSense, and HAPT. In addition to comparing our self-supervised model with frozen convolution blocks to other works, to evaluate the impact of fine-tuning on our self-supervised model, we further re-train $H_{HAR}$.
Tables \ref{tab:UCI_score}, \ref{tab:MotionSenhse_score}, and \ref{tab:HAPT_score} show that for the UCI HAR dataset, fine-tuning does not help the performance, while a small improvement is achieved for both MotionSense and HAPT datasets. This indicates that the original representations learned by our self-supervised network are effective to a point where further training, even with labeled data, does not help. Nonetheless, for both UCI HAR and HAPT, our fine-tuned self-supervised models outperform the fine-tuned models proposed by \cite{saeed2019multi} and \cite{saeed2020sense}.

\begin{figure}[t!]
	\centering
	\includegraphics[width=1\columnwidth]{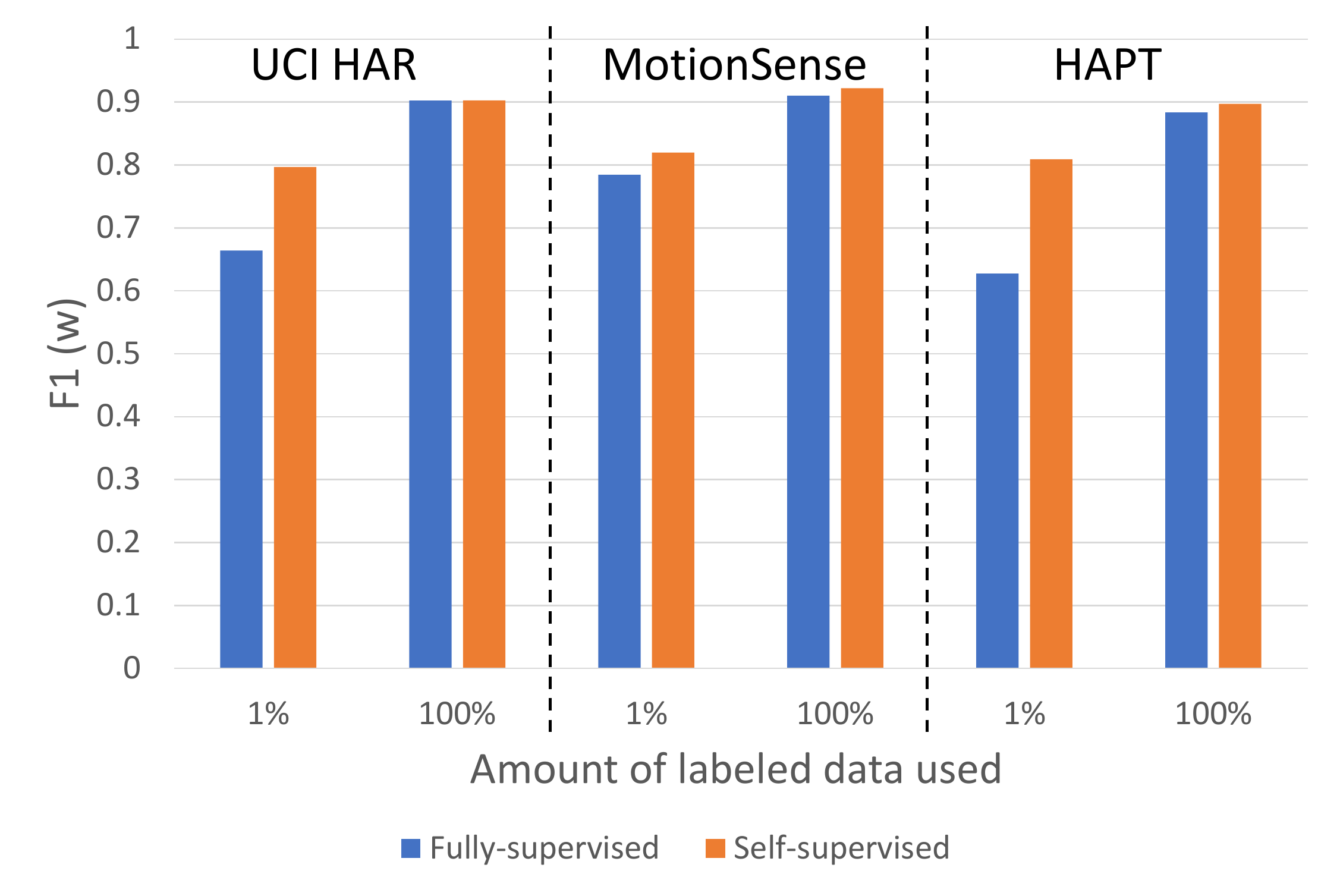}
	\caption{Performance of the proposed method on UCI HAR, MotionSense, and HAPT datasets with the self-supervised and fully-supervised approaches when different amounts of labeled data (1\% and 100\%) are available.}
	\label{fig:labeled data}
\end{figure}

Moreover, we evaluate the robustness of our self-supervised model in comparison to a fully-supervised one with the same architecture, when different amounts of labeled data are used. For the self-supervised model, we first train the pretext motion prediction model using the described self-supervised scheme. Next, we perform weight transfer to the HAR network with the same architecture and use either 100\% of the labeled data or only 1\% of the labeled data for training the FC layers. For the fully supervised experiment, however, we simply train the HAR network end-to-end with either all or only 1\% (the same instances of data used to train the downstream task in the self-supervised experiment) of the labeled data. The results of these experiments are presented in Figure \ref{fig:labeled data}. We observe that the self-supervised and fully-supervised methods achieve relatively similar results when 100\% of the labeled data re used. However, when only smaller amounts (1\%) of labeled data are available, the self-supervised model clearly outperforms the fully-supervised one, demonstrating that our proposed method can learn to extract effective representations with small amounts of labeled data.

\begin{table}[!t]
\centerline{}
\caption{UCI HAR dataset results. SS: self-supervised, FT: fine-tuned.}
\small
\setlength
\tabcolsep{2pt}
\begin{tabular}{llll}
\hline
\textbf{Method} & \textbf{Accuracy} & \textbf{F1(m)} & \textbf{F1(w)} \\ \hline\hline
Saeed et al. \cite{saeed2019multi} (SS)                & --           & --           & 0.890$\pm$0.044               \\
Haresamudram et al. \cite{haresamudram2020masked} (SS) &--    & 0.819   & --   \\ 
\textbf{Ours} (SS)  &\textbf{0.908}$\pm$\textbf{0.036}           & \textbf{0.910}$\pm$\textbf{0.034}         & \textbf{0.908}$\pm$\textbf{0.036}   \\ \hline
Saeed et al. \cite{saeed2019multi} (SS, FT)           & --           & --          & 0.905$\pm$0.050                 \\
\textbf{Ours} (SS, FT) &\textbf{0.906}$\pm$\textbf{0.038}         & \textbf{0.908}$\pm$\textbf{0.036}     & \textbf{0.906}$\pm$\textbf{0.038}   \\ \hline  
\end{tabular}
\label{tab:UCI_score}
\end{table}

\begin{table}[!t]
\centerline{}
\caption{MotionSense dataset results. SS: self-supervised, FT: fine-tuned.}
\small
\setlength
\tabcolsep{2pt}
\begin{tabular}{llll}
\hline
\textbf{Method} & \textbf{Accuracy} & \textbf{F1(m)} & \textbf{F1(w)} \\ \hline\hline
Saeed et al. \cite{saeed2019multi} (SS)                & --           & --           & 0.919$\pm$0.019                \\
Haresamudram et al. \cite{haresamudram2020masked} (SS) &--    & 0.880   & --   \\
\textbf{Ours} (SS)  &\textbf{0.920}$\pm$\textbf{0.022}           & \textbf{0.901}$\pm$\textbf{0.027}         & \textbf{0.921}$\pm$\textbf{0.022}   \\ \hline
Saeed et al. \cite{saeed2019multi} (SS, FT)           & --           & --          & \textbf{0.939}$\pm$\textbf{0.025}      \\
\textbf{Ours} (SS, FT) &\textbf{0.933}$\pm$\textbf{0.024}  & \textbf{0.918}$\pm$\textbf{0.029}     & 0.934$\pm$0.023   \\ \hline  \end{tabular}
\label{tab:MotionSenhse_score}
\end{table}

\begin{table}[!t]
\centerline{}
\caption{HAPT dataset results. SS: self-supervised, FT: fine-tuned.}
\small
\setlength
\tabcolsep{2pt}
\begin{tabular}{llll}
\hline
\textbf{Method} & \textbf{Accuracy} & \textbf{F1(m)} & \textbf{F1(w)} \\ \hline\hline
Saeed et al. \cite{saeed2020sense} (SS)                & --           & --           & 0.863$\pm$0.045                \\
\textbf{Ours} (SS)  &\textbf{0.899}$\pm$\textbf{0.034}           & \textbf{0.796}$\pm$\textbf{0.025}         & \textbf{0.898}$\pm$\textbf{0.034}   \\ \hline
Saeed et al. \cite{saeed2020sense} (SS, FT)           & --           & --           &  0.896$\pm$0.051                 \\
\textbf{Ours} (SS, FT) &\textbf{0.901}$\pm$\textbf{0.038}         & \textbf{0.792}$\pm$\textbf{0.029}     & \textbf{0.900}$\pm$\textbf{0.038}   \\ \hline  
\end{tabular}
\label{tab:HAPT_score}
\end{table}

\section{Conclusion and Future Work}
\label{ssec:Conclusion and Future Work}
In this study, we proposed a novel deep self-supervised solution for learning representations from unlabeled accelerometer data. A deep neural network was first trained on raw signals, learning to predict the values in the \textit{z} dimension of accelerometer data,  utilizing the past values of that intended dimension as well as both past and present motion in the other two dimensions. This pre-trained model was then frozen and transferred to a supervised human activity recognition network. 
We show that our method outperforms existing self-supervised methods in HAR for UCI HAR, MotionSense, and HAPT datasets.

For future work, using additional datasets for the pretext self-supervised task could potentially improve the model performance in extracting more generalized motion representations from accelerometer data. Furthermore, using CNN-RNN-based methods may help the model in better learning the temporal dependencies of the timeseries accelerometer data.

\vspace*{1cm}

\bibliographystyle{ACM-Reference-Format}
\bibliography{sample-base}

\end{document}